\documentclass[12pt]{article}
\usepackage{ctex}
\usepackage{amssymb,amsmath,color,amsmath,bm}
\usepackage[colorlinks,linkcolor=blue]{hyperref}
\usepackage{geometry}
\usepackage{tabularx}
\usepackage{bm}
\usepackage{authblk}
\usepackage[square, comma, sort&compress, numbers]{natbib}

	\title{\bf Self-dual twisted generalized Reed-Solomon codes}	
		\author{\small Canze Zhu$^{1}$}
		\author{\small Qunying Liao$^{2}$
			\thanks{{Corresponding author$^{2}$. E-mail: qunyingliao@sicnu.edu.cn;
					
					{\quad Contributing author$^{1}$. E-mail: ~canzezhu@163.com.}}	
		}}
		\affil[]{\small (College of Mathematical Science, Sichuan Normal University, Chengdu Sichuan, 610066)}
		\date{}
	\newtheorem{theorem}{Theorem}[section]
	\newtheorem{definition}{Definition}[section]
	\newtheorem{lemma}{Lemma}[section]

	\newtheorem{corollary}{Corollary}[section]
	\newtheorem{remark}{Remark}[section]

	\catcode`@=11 \@addtoreset{equation}{section} \catcode`@=12
\begin{document}
	\maketitle
	{\bf Abstract.}
	{\small In this paper, by using some properties for linear algebra methods, the parity-check matrices for twisted generalized Reed-Solomon codes with any given hook $h$ and twist $t$ are presented, and then a sufficient and necessary condition for the twisted generalized Reed-Solomon code with $h\ge t$ to be self-dual is given. Furthermore, several classes of self-dual codes with small Singleton defect are constructed based on twisted generalized Reed-Solomon codes, especially some of these self-dual codes are  MDS or NMDS.
	}\\
	
	
	\section{Introduction}
	In this paper, let $\mathbb{F}_{q}$ be the finite field with $q$ elements and $\mathbb{F}_{q}^*=\mathbb{F}_{q}\backslash \{0\}$, where $q$ is a prime power. An $[n,k,d]$ linear code $\mathcal{C}$ over $\mathbb{F}_q$ is a $k$-dimensional subspace of $\mathbb{F}_q^n$ with minimum (Hamming) distance $d$ and length $n$. For any $\mathbf{a}=(a_1,\ldots,a_n)$, $\mathbf{b}=(b_1,\ldots,b_n)\in\mathbb{F}_q^n$, the inner product is defined as $\langle \mathbf{a},\mathbf{b}\rangle=\sum\limits_{i=1}^{n}a_ib_i$, and 
	the dual code of $\mathcal{C}$ is defined as $$\mathcal{C}^{\perp}=\{\mathbf{c}^{'}\in\mathbb{F}_q^n~|~\langle\mathbf{c}^{'},\mathbf{c}\rangle=0, \text{for any}~\mathbf{c}\in\mathcal{C}\}.$$ $\mathcal{C}$ is called self-dual code if $\mathcal{C}=\mathcal{C}^{\perp}$.
	In recent years, study of self-dual maximum distance separable (in short, MDS) codes have attracted a lot of attention \cite{1,4,11,12,13,14,15,20,21,33,34,37}. On the one hand, due to their nice structures, self-dual codes have attracted attention by coding theorists, cryptographers and mathematicians, or rather, self-dual codes have
	various applications in cryptography (in particular secret sharing) \cite{5,9,29} and combinatorics \cite{17,28}. On the other hand, the MDS code not only 
	can correct maximal number of errors for a given code rate, but also are closely connected to combinatorial designs and finite geometry \cite{17,28}. Thus, it is natural to consider the intersection of these two classes of codes, namely, self-dual MDS codes. Especially, generalized Reed-Solomon (in short, GRS) codes are a class of MDS codes and a lot of self-dual MDS codes from GRS codes are constructed in \cite{10,19,25,39,40}.
	
	For a non-negative integer $m$, the $m$-MDS code is introduced independently in \cite{27, 41}. A linear code $\mathcal{C}$ is called an $m$-MDS code if both $\mathcal{C}$ and its dual code have $m$ Singleton defect from being an MDS code. Especially, for $m=0$ or $1$,  $\mathcal{C}$ is called MDS or Near MDS (NMDS) \cite{7}, respectively. It is easy to see that a self-dual code is an $m$-MDS code. For small $m$, $m$-MDS codes approximate maximal minimum Hamming distance for a given code rate. Furthermore, $m$-MDS codes also have been applied in secret sharing scheme \cite{35}, the index coding problem \cite{42} and the informed source coding problem \cite{43}. From both theoretical and practical points of view, to study self-dual $m$-MDS codes with small $m$ is interesting. 
	
	In 2017,  inspired by twisted Gabidulin codes \cite{t26}, twisted Reed-Solomon (in short, TRS) codes were firstly introduced by Beelen et al. as a generalization for Reed-Solomon codes, they also showed that TRS codes could be well decoded. Furthermore, different from GRS codes, the twisted generalized Reed-Solomon  (in short, TGRS) code is not necessarily MDS \cite{2}. The authors also showed that most of TGRS MDS codes are not GRS codes \cite{3,26}. Furthermore, basing on TGRS codes, Lavauzelle et al. presented an efficient key-recovery attack used in the McEliece cryptosystem \cite{24}. TGRS codes are also used to construct linear complementary dual MDS codes by their applications in cryptography \cite{16,26}. Recently, Huang et, al. gave the parity-check matrix for a TGRS code with  hook $k-1$ and twist $1$, and then constructed several classes of self-dual MDS or NMDS codes from these TGRS codes \cite{t0}.
	
	In this paper, we extend the main results in \cite{t0}. In Section 2, some basic notations and results about TGRS codes are given. In Section 3,  the parity-check matrix for the TGRS code with any given hook $h$ and twist $t$ is presented. In Section 4, a sufficient and necessary condition for the TGRS code with $h\ge t$ to be self-dual is given, and then several classes of self-dual $m$-MDS codes with $m\le t$ are constructed. In Section 5, we conclude the whole paper.
	
	\section{Preliminaries}
	
	In this section, we mainly give some notations and lemmas about  $m$-MDS codes and TGRS codes.
	
	For an $[n,k,d]$ linear code $\mathcal{C}$, the singleton bound implies that $n+1-k-d\ge 0$, inspired by which,  for a non-negative integer $m$, the $m$-MDS code is defined as follows.
	\begin{definition} [\cite{7, 27,40}]
		For a linear code $\mathcal{C}$ with parameters $[n,k,d]$, the Singleton defect is defined as $S(\mathcal{C})=n+1-k-d$. $\mathcal{C}$ is called an $m$-MDS code if $S(\mathcal{C})=S(\mathcal{C^{\perp}})=m$. Especially, $\mathcal{C}$ is called an MDS or  NMDS code if $m=0$ or $1$, respectively. 
	\end{definition}
	
	Some related definitions of TGRS codes are given in the following. 
	\begin{definition}[\cite{2,3,24}]\label{d23} 
		Let $t$, $h$ and $k$ be integers with $0\le h<k\le q$ and $\eta\in\mathbb{F}_q^{*}$. Define the set of $(k,t,h,\eta)$-twisted polynomials as 
		\begin{align*}
			\mathcal{V}_{k,t,h,\eta}=\Big\{f(x)=\sum_{i=0}^{k-1}a_ix^i+\eta a_hx^{k-1+t}~|~a_i\in\mathbb{F}_q~(i=0,\ldots,k-1)\Big\},
		\end{align*}
		which is a $k$-dimensional $\mathbb{F}_q$-linear subspace. We call $h$ the hook and $t$ the twist.
	\end{definition}
	
	\begin{definition} [\cite{2,3,24}]
		Let $t$, $h$, $k$ and $n$ be positive integers and  $\eta\in\mathbb{F}_q^{*}$. Let $\boldsymbol{\alpha}=(\alpha_1,\ldots,\alpha_n)\in\mathbb{F}_q^n$ with $\alpha_i\neq \alpha_j$ $(i\neq j)$ and
		$\boldsymbol{v} = (v_1,\ldots,v_n)\in (\mathbb{F}_q^{*})^n$. The TGRS code with length $n$ and dimension $k$ is defined as
		\begin{align*}
			\mathcal{C}_{k,n}(\boldsymbol{\alpha},\boldsymbol{v},t,h,\eta)=\{(v_1f({\alpha_{1}}),\ldots,v_nf(\alpha_{n}))~|~f(x)\in  \mathcal{V}_{k,t,h,\eta}\}.
		\end{align*}	
		Especially,	for $\texorpdfstring{\bm{1}}~=(1,\ldots,1)$,  $\mathcal{C}_{k,n}(\texorpdfstring{\bm{\alpha}}~,\texorpdfstring{\bm{1}}~,t,h,\eta)$ is called a TRS code.
	\end{definition}
	
	By Definition \ref{d23}, it is easy to see that the generator matrix of $\mathcal{C}_{k,n}(\boldsymbol{\alpha},\boldsymbol{v},t,h,\eta)$ is
	\begin{align*}
	G_k(\boldsymbol{v})=\left(\begin{matrix}
	&v_1~&v_2~&\ldots~&v_n\\
	&v_1\alpha_1~&v_2\alpha_2~&\ldots~&v_n\alpha_{n}\\
	&\vdots~&\vdots&~&\vdots\\
	&v_1\big(\alpha_1^h+\eta\alpha_1^{k-1+t}\big)~&v_2\big(\alpha_2^h+\eta\alpha_2^{k-1+t}\big)~&\ldots~&v_n\big(\alpha_{n}^h+\eta\alpha_n^{k-1+t}\big)\\
	&\vdots~&\vdots&~&\vdots\\
	&v_1\alpha_1^{k-1}~&v_2\alpha_{2}^{k-1}~&\cdots~&v_n\alpha_{n}^{k-1}
	\end{matrix}\right).
	\end{align*}
	
	The following lemma gives a sufficient condition for a TGRS code to be MDS.
	\begin{lemma}[Theorem 17, in \cite{2}] \label{lmds}
		Let $\mathbb{F}_s$ be a proper subfield of $\mathbb{F}_q$ and $\texorpdfstring{\bm{\alpha}} ~\in\mathbb{F}_s^{n}$. If $\eta\in \mathbb{F}_q\backslash \mathbb{F}_s$, then  $\mathcal{C}_{k,n}(\texorpdfstring{\bm{\alpha}}~,\texorpdfstring{\bm{v}}~,t,h,\eta)$ is MDS.
	\end{lemma}

	The following lemma gives the bound for the minimum distance of $\mathcal{C}_{k,n}^{\perp}(\boldsymbol{\alpha},\boldsymbol{v},t,h,\eta)$.
	\begin{lemma}\label{0h}
		$\mathcal{C}_{k,n}^{\perp}(\boldsymbol{\alpha},\boldsymbol{v},t,h,\eta)$ is an $[n,n-k,d]$ linear code with $h+1\le d\le k+1$.	
	\end{lemma}

	{\bf Proof}.~If $d\le h$, then there are $1\le i_1<\cdots<i_{d}\le n$, such that
	\begin{align*}
	D(i_1,\ldots,i_d)=\left(\begin{matrix}
	&1~&1~&\ldots~&1\\
	&\vdots~&\vdots&~&\vdots\\
	&\alpha_{i_1}^h+\eta\alpha_{i_1}^{k-1+t}~&\alpha_{i_2}^h+\eta\alpha_{i_2}^{k-1+t}~&\ldots~&\alpha_{i_{d}}^h+\eta\alpha_{i_{d}}^{k-1+t}\\
	&\vdots~&\vdots&~&\vdots\\
	&\alpha_{i_1}^{k-1}~&\alpha_{i_2}^{k-1}~&\cdots~&\alpha_{i_{d}}^{k-1}
	\end{matrix}\right),
	\end{align*}
	with $d$ linearly dependent columns, thus $\mathrm{Rank}\big(D(i_1,\ldots,i_d)\big)<d$, which leads $\mathbf{1}$, $\tilde{\boldsymbol{\alpha}}$,$\ldots$,$\tilde{\boldsymbol{\alpha}}^{d-1}$ are linearly dependent, where 
	\begin{align*}
	\tilde{\boldsymbol{\alpha}}^{j}=(\alpha_{i_1}^{j},\alpha_{i_2}^{j},\ldots,\alpha_{i_{d}}^{j})\quad (j=1,\ldots,d-1).
	\end{align*}Thus there exist not all zero elements $b_{0},b_{1},\ldots,b_{d-1}$ in $\mathbb{F}_q$, such that $\alpha_{i_1},\ldots,\alpha_{i_{d}}$ are roots of the polynomial $f(x)=\sum\limits_{i=0}^{d-1}b_ix^i$, it contradicts with  $\deg f(x)\le d-1$. Thus $d\ge h+1$. Furthermore, the upper bound of $d$ follows from the Singleton bound directly.
	$\hfill\Box$\\
 	
 	It is easy to see that a self-dual code is always an $m$-MDS code, the following lemma gives the  upper bound for the Singleton defect of the self-dual TGRS code. 
 	\begin{lemma}\label{lh}
 		If $\mathcal{C}_{k,n}(\boldsymbol{\alpha},\boldsymbol{v},t,h,\eta)$ is self-dual, then it is an $m$-MDS code with $$m\le\min\{t,k-h\}.$$
 	\end{lemma}
 
 	{\bf Proof}. By Lemma $\ref{0h}$, we have $$S(\mathcal{C}_{k,n}^{\perp}(\boldsymbol{\alpha},\boldsymbol{v},t,h,\eta))\le k+1-(h+1)=k-h.$$ Furthermore, by the definition of the TGRS code, one has  $$S(\mathcal{C}_{k,n}(\boldsymbol{\alpha},\boldsymbol{v},t,h,\eta))\le t.$$  Thus if $\mathcal{C}_{k,n}(\boldsymbol{\alpha},\boldsymbol{v},t,h,\eta)$ is  self-dual, then 
 	\begin{align*}
 	\qquad\qquad	S(\mathcal{C}_{k,n}(\boldsymbol{\alpha},\boldsymbol{v},t,h,\eta))=S(\mathcal{C}_{k,n}^{\perp}(\boldsymbol{\alpha},\boldsymbol{v},t,h,\eta))\le\min\{t,k-h\}.\qquad\qquad\hfill\Box
 	\end{align*}
 	
 	The following lemma gives a sufficient and necessary condition for $\mathcal{C}_{k,n}(\boldsymbol{\alpha},\texorpdfstring{\bm{v}}~,1,k-1,\eta)$ to be MDS or NMDS, respectively.
	\begin{lemma}[Lemma 2.6 in \cite{t0}]\label{l}For positive integers $k$ and $n$ with $k<n$ and  $\boldsymbol{\alpha}=(\alpha_1,\ldots,\alpha_n)\in\mathbb{F}_q^{n}$ with $\alpha_i\neq\alpha_j$ $(i\neq j)$, denote $$S_k(1,\boldsymbol{\alpha})=\Big\{\sum_{i\in I}\alpha_i~\bigg|~\forall I\subsetneq \{1,\ldots,n\} \text{~with~}|I|=k\Big\},$$ then we have
		
 $(1)$	$\mathcal{C}_{k,n}(\texorpdfstring{\bm{\alpha}}~,\texorpdfstring{\bm{v}}~,1,k-1,\eta)$ is an MDS code if and only if $\eta^{-1}\notin S_k(1,\boldsymbol{\alpha})$;
		
 $(2)$	$\mathcal{C}_{k,n}(\texorpdfstring{\bm{\alpha}}~,\texorpdfstring{\bm{v}}~,1,k-1,\eta)$ is an NMDS code if and only if $\eta^{-1}\in S_k(1,\boldsymbol{\alpha})$.
		
	\end{lemma}

	In the following lemma, a sufficient and necessary condition for the Singleton defect of $\mathcal{C}_{k,n}(\texorpdfstring{\bm{\alpha}}~,\texorpdfstring{\bm{v}}~,2,k-2,\eta)$ equal to $0,1$$\text{~or~}$$2$ is given, respectively. 
	 \begin{lemma}\label{l2}
		For positive integers $k$ and $n$ with $k<n$ and  $\boldsymbol{\alpha}=(\alpha_1,\ldots,\alpha_n)\in\mathbb{F}_q^{n}$ with $\alpha_i\neq\alpha_j$ $(i\neq j)$, denote 
		{\small
		\begin{align*}
		S_k(2,\boldsymbol{\alpha})=\left\{\sum_{i\in I}\alpha_i\sum_{\substack{{i,j\in I}\\i\neq j}}\alpha_{i}\alpha_{j}-\!\!\!\!\sum_{\substack{{i,j,l\in I}\\{(i-j)(i-l)(j-l)\neq 0}}}\!\!\!\!\alpha_{i}\alpha_{j}\alpha_{l}~\bigg|~\forall I\subsetneq \{1,\ldots,n\} \text{~with~}|I|=k\right\}
		\end{align*}}
	and
	{\small\begin{align*}
			\tilde{S}_k(2,\boldsymbol{\alpha})=\left\{-\!\!\!\!\sum_{\substack{{i,j,l\in \tilde{I}}\\{(i-j)(i-l)(j-l)\neq 0}}}\!\!\!\!\alpha_{i}\alpha_{j}\alpha_{l}~\bigg|~\forall \tilde{I}\subseteq \{1,\ldots,n\} \text{~with~}|\tilde{I}|=k+1~\text{and}~\sum\limits_{i\in\tilde{I}}\alpha_i=0\right\},
		\end{align*}
	}
then we have

	  $(1)$ $\mathcal{C}_{k,n}(\texorpdfstring{\bm{\alpha}}~,\texorpdfstring{\bm{v}}~,2,k-2,\eta)$ is an MDS code if and only if $\eta^{-1}\notin S_k(2,\boldsymbol{\alpha})$;
		
	$(2)$ $S\big(\mathcal{C}_{k,n}(\texorpdfstring{\bm{\alpha}}~,\texorpdfstring{\bm{v}}~,2,k-2,\eta)\big)=1$ if and only if $\eta^{-1}\in S_k(2,\boldsymbol{\alpha})\backslash \tilde{S}_k(2,\boldsymbol{\alpha})$;
		
	$(3)$ $S\big(\mathcal{C}_{k,n}(\texorpdfstring{\bm{\alpha}}~,\texorpdfstring{\bm{v}}~,2,k-2,\eta)\big)=2$ if and only if $\eta^{-1}\in S_k(2,\boldsymbol{\alpha})\cap\tilde{S}_k(2,\boldsymbol{\alpha})$.
	\end{lemma}

	{\bf  The proof for Lemma \ref{l2}}. By the definition of the TGRS code, it is easy to see that $S\big(\mathcal{C}_{k,n}(\boldsymbol{\alpha},\texorpdfstring{\bm{v}}~,2,k-2,\eta)\big)\le 2$. 
	In the following, we firstly give the proof for $(1)$, and then show that $(3)$ holds, finally, by $(1)$ and $(3)$, we can get $(2)$ directly.
	
 For the convenience, we denote $w_H(\mathbf{c}_f)$ to be the Hamming weight for a codeword $\mathbf{c}_f=\big(f(\alpha_1),\ldots,f(\alpha_n)\big)\in\mathcal{C}_{k,n}(\texorpdfstring{\bm{\alpha}}~,\texorpdfstring{\bm{v}}~,2,k-2,\eta)$.
	 
{\bf The proof for $(1)$}. It is enough to prove that the converse-negative proposition for $(1)$ holds. In fact, if $S\big(\mathcal{C}_{k,n}(\boldsymbol{\alpha},\texorpdfstring{\bm{v}}~,2,k-2,\eta)\big)\neq 0$, then there exists a $I\subsetneq \{1,\ldots,n\}$ with $|I|=k$ and $f(x)=a_0+a_1x+\cdots+a_{k-1}x^{k-1}+\eta a_{k-2} x^{k+1}\in\mathbb{F}_{q}[x]\backslash\{0\}$, such that $f(\alpha_i)=0~(\forall i\in I)$,
	which implies $a_{k-2}\neq 0$. Thus there exists some $b\in\mathbb{F}_q$ such that
	\begin{align*}
	f(x)=\eta a_{k-2}(x-b)\prod_{i\in I}(x-\alpha_i). 
	\end{align*}
	It leads that
	\begin{align*}
	-b-\sum_{i\in I}\alpha_i=0\quad\text{and}\quad-\eta\left(b\sum_{\substack{{i,j\in I}\\i\neq j}}\alpha_{i}\alpha_{j}+\sum_{\substack{{i,j,l\in I}\\{(i-j)(i-l)(j-l)\neq 0}}}\alpha_{i}\alpha_{j}\alpha_{l}\right)=1,
	\end{align*}
	and then $\eta^{-1}\in S_k(2,\boldsymbol{\alpha})$.
	
	Conversely, for $\eta^{-1}\in S_k(2,\boldsymbol{\alpha})$, there exists some $I\subsetneq\{1,\ldots,n\}$ with $|I|=k$, such that
	\begin{align*}
	\eta= \left(\sum_{i\in I}\alpha_i\sum_{\substack{{i,j\in I}\\i\neq j}}\alpha_{i}\alpha_{j}-\sum_{\substack{{i,j,l\in I}\\{(i-j)(i-l)(j-l)\neq 0}}}\alpha_{i}\alpha_{j}\alpha_{l}\right)^{-1},
	\end{align*}
	thus $f(x)=\eta(x+\sum\limits_{i\in I}\alpha_i)\prod\limits_{i\in I}(x-\alpha_i)\in \mathcal{V}_{k,t,h,\eta}$ and 
	$w_H(\boldsymbol{c}_f)\le n-k$, which means that $$S\big(\mathcal{C}_{k,n}(\boldsymbol{\alpha},\texorpdfstring{\bm{v}}~,2,k-2,\eta)\big)\neq 0.$$
	 
	By the above discussions, $S\big(\mathcal{C}_{k,n}(\boldsymbol{\alpha},\texorpdfstring{\bm{v}}~,2,k-2,\eta)\big)=0$ if and only if $\eta^{-1}\notin S_k(2,\boldsymbol{\alpha})$.

{\bf The proof for $(3)$}. If $S(\mathcal{C}_{k,n}(\boldsymbol{\alpha},\texorpdfstring{\bm{v}}~,2,k-2,\eta))=2$, then there exist some $\tilde{I}\subseteq \{1,\ldots,n\}$ with $|\tilde{I}|=k+1$ and $f(x)=a_0+a_1x+\cdots+a_{k-1}x^{k-1}+\eta a_{k-2} x^{k+1}\in\mathbb{F}_{q}[x]\backslash\{0\}$, such that $f(\alpha_i)=0~ (\forall i\in \tilde{I})$,
	which implies $a_{k-2}\neq 0$, thus
	\begin{align*}
	f(x)=\eta a_{k-2}\prod_{i\in \tilde{I}}(x-\alpha_i),
	\end{align*}
	and then
	\begin{align*}
	\eta^{-1}=-\sum_{\substack{{i,j,l\in \tilde{I}}\\{(i-j)(i-l)(j-l)\neq 0}}}\alpha_{i}\alpha_{j}\alpha_{l}.
	\end{align*}
	Conversely, for $\eta^{-1}\in \tilde{S}_k(2,\boldsymbol{\alpha})$, there exist some $\tilde{I}\subseteq\{1,\ldots,n\}$ with $|\tilde{I}|=k+1$, such that
	\begin{align*}
	\eta= \left(-\sum_{\substack{{i,j,l\in \tilde{I}}\\{(i-j)(i-l)(j-l)\neq 0}}}\!\!\!\!\alpha_{i}\alpha_{j}\alpha_{l}\right)^{-1},
	\end{align*}
	thus $f(x)=\eta\prod\limits_{i\in \tilde{I}}(x-\alpha_i)\in \mathcal{V}_{k,t,h,\eta}$ and 
	$w_H(\boldsymbol{c}_f)\le n-k-1$, which leads $$S\big(\mathcal{C}_{k,n}(\boldsymbol{\alpha},\texorpdfstring{\bm{v}}~,2,k-2,\eta)\big)=2.$$
	So far, we know that $S\big(\mathcal{C}_{k,n}(\texorpdfstring{\bm{\alpha}}~,\texorpdfstring{\bm{v}}~,2,k-2,\eta)\big)=2$ if and only if $$\eta^{-1}\in S_k(2,\boldsymbol{\alpha})\cap\tilde{S}_k(2,\boldsymbol{\alpha}).$$
	
{\bf The proof for $(2)$}.	By $(1)$ and $(3)$, we get $(2)$ directly.	$\hfill\Box$\\

				
	
	

\section{The parity-check matrix for the TGRS code}

By the theory of Vandermonde matrices \cite{30},  one has the following lemma, which is necessary to calculate the parity-check matrix for the code $\mathcal{C}_{k,n}(\texorpdfstring{\bm{\alpha}}~,\texorpdfstring{\bm{v}}~,t,h,\eta)$.
\begin{lemma}\label{l1}
	Denote 
	\begin{align*}
	G=\left(\begin{matrix}
	&1~&1~&\ldots~&1\\
	&\alpha_1~&\alpha_2~&\ldots~&\alpha_{n}\\
	&\vdots~&\vdots&~&\vdots\\
	&\alpha_1^{n-1}~&\alpha_{2}^{n-1}~&\cdots~&\alpha_{n}^{n-1}
	\end{matrix}\right),
	\end{align*}	then for the system of equations over $\mathbb{F}_q$
	\begin{align}\label{GG}
	G(u_1,u_2,\ldots,u_n)^T=(0,0,\ldots,1)^T,
	\end{align} there is an unique solution $(u_1,u_2,\ldots,u_n)^T$, where $u_i=\prod\limits_{j=1,j\neq i}^{n}(\alpha_i-\alpha_{j})^{-1}$ $(1\le i\le n)$. 
\end{lemma}

The parity-check matrix for $\mathcal{C}_{k,n}(\texorpdfstring{\bm{\alpha}}~,\texorpdfstring{\bm{v}}~,t,h,\eta)$ is given in Theorem \ref{31}. In order to give the proof, we need the following nonations.  

For $r\in\mathbb{F}_q$, $1\le i_1,i_2\le n$ and  $n\times n$ matrix $M$, let $P\big(i_1,i_2(r)\big)$, $Q\big(i_1,i_2(r)\big)$ and $T\big(i_1,i_2\big)$ be $n\times n$ elementary matrices satisfying the following conditions, respectively.

$\bullet$ the $i_1$th row of $P\big(i_1,i_2(r)\big)M$ is replaced by the sum of $r$ times the $i_2$th row and  the $i_1$th row of $M$;

$\bullet$ the $i_1$th column of $MQ\big(i_1,i_2(r)\big)$ is replaced by the sum of $r$ times the $i_2$th column and  the $i_1$th column of $M$;

$\bullet$ $MT(i_1,i_2)$ that exchanges the $i_1$th column and the $i_2$th column of $M$.\\

\begin{theorem}\label{31}
	For any integer $m$, let $L_m=\sum\limits_{l=1}^{n}u_l\alpha_{l}^{n-1+m}$ and $u_i=\prod\limits_{j=1,j\neq i}^{n}(\alpha_i-\alpha_{j})^{-1}$ $(i=1,\ldots,n)$, then $\mathcal{C}_{k,n}(\texorpdfstring{\bm{\alpha}}~,\texorpdfstring{\bm{v}}~,t,h,\eta)$ has the parity-check matrix  $$H_{n-k}(\boldsymbol{v})=\left(\begin{matrix}\boldsymbol{\beta}_1,\boldsymbol{\beta}_2,\ldots,\boldsymbol{\beta}_n
	\end{matrix}\right),$$ 
	where 
	\begin{align*}
	\boldsymbol{\beta}_{j}
	=\left(\begin{matrix}
	&\frac{u_j}{v_j}\\
	&\frac{u_j}{v_j}\alpha_j\\
	&\vdots\\
	&\frac{u_j}{v_j}\alpha_j^{n-(k+t+1)}\\
	&\frac{u_j}{v_j}\Big(\alpha_j^{n-(h+1)}-\sum\limits_{m=1}^{k-h-1}L_{m}\alpha_j^{n-(h+1)-m}+\tilde{L}\alpha_j^{n-(k+t)}\Big)\\
	&\frac{u_j}{v_j}\big(\alpha_j^{n-(k+t-1)}-L_1\alpha_j^{n-(k+t)}\big)\\
	&\vdots\\
	&\frac{u_j}{v_j}\big(\alpha_j^{n-(k+1)}- L_{t-1}\alpha_j^{n-(k+t)}\big)\\
	\end{matrix}\right)_{(n-k)\times 1}
	\end{align*}	
	and\begin{align*}
	\tilde{L}=\sum\limits_{m=1}^{k-h-1}L_{m} L_{k+t-h-1-m}-\eta^{-1}(1+\eta L_{k+t-h-1}).
	\end{align*}	
\end{theorem}

{\bf {Proof}}. It is easy to see that
\begin{align}\label{HH1}
	 H_{n-k}(\boldsymbol{v})=\left(\begin{matrix}
	&v_1^{-1}~&0~&\ldots~&0\\
	&0~&v_2^{-1}~&\ldots~&0\\
	&\vdots~&\vdots~&~&\vdots\\
	&0~&0~&\ldots~&v_n^{-1}\\
	\end{matrix}\right)H_{n-k}(\mathbf{1}).
\end{align}
 Thus it is enough to investigate the parity-check matrix $H_{n-k}(\mathbf{1})$ for $\mathcal{C}_{k,n}(\texorpdfstring{\bm{\alpha}}~,\texorpdfstring{\bm{1}}~,t,h,\eta)$.
Let
\begin{align*}
G=\left(\begin{matrix}
&1~&1~&\ldots~&1\\
&\alpha_1~&\alpha_2~&\ldots~&\alpha_{n}\\
&\vdots~&\vdots&~&\vdots\\
&\alpha_1^{n-1}~&\alpha_{2}^{n-1}~&\cdots~&\alpha_{n}^{n-1}
\end{matrix}\right)
\end{align*}
and
\begin{align*}
H=\left(\begin{matrix}
&u_1\alpha_1^{n-1}~&\ldots~&u_1\alpha_{1}^{n-k-1}~&\ldots~&u_1\\
&u_2\alpha_2^{n-1}~&\ldots~&u_2\alpha_{2}^{n-k-1}~&\ldots~&u_2\\
&\vdots~&~&\vdots~&~&\vdots\\
&u_n\alpha_n^{n-1}~&\ldots~&u_n\alpha_{n}^{n-k-1}~&\ldots~&u_n\\
\end{matrix}\right),
\end{align*}then one has
\begin{align*}
GH=(l_{i,j})_{1\le i,j\le n}
\end{align*}with $l_{i,j}=L_{i-j}=\sum\limits_{l=1}^{n}u_l\alpha_{l}^{n-1+i-j}$.  By Lemma \ref{l1}, we can get
\begin{align*}
l_{i,j}=\begin{cases}
0,\quad&i<j;\\
1,\quad&i=j.\\
\end{cases}
\end{align*}

For the matrix $GH$, replacing the $(h+1)$th row by the sum of $\eta$ times $(k+t)$th row and  $(h+1)$th row, one has
\begin{align*}	&P\big(h+1,(k+t)(\eta)\big)GH\\
&\quad\begin{array}{ccccccccccc} 1 & \qquad &~\quad h&~\quad h+1&\qquad\quad&~\quad k+1&\qquad&~~~k+t& & &\end{array}\\
= &\left( 
\begin{array}{ccccccccccc} 	
1      &\ldots&  0    &0         & \cdots       & 0 &\ldots   &0               &0     &\cdots &0 \\
\vdots &      & \vdots&\vdots  &          &\vdots&      &\vdots          &\vdots&       &\vdots \\
*      & \cdots      & *     &1+\eta l_{k+t,h+1}& \cdots &\eta l_{k+t,k+1}&\ldots&\eta &0     &\cdots &0\\
\vdots &      & \vdots&\vdots            &&   \vdots   &         &\vdots&\vdots   &  &\vdots \\	 *      & \cdots      & *     &*              & \cdots  &0               & \cdots     &0          &0        & \cdots      &0 \\
*      & \cdots      & *     &*              & \cdots    &1               & \cdots     &0          &0        & \cdots      &0 \\
\vdots &      & \vdots&\vdots            && \vdots     &        &\vdots& \vdots      &&\vdots \\
*      &  \cdots     & *     &*         & \cdots         &*     &\cdots      &1               &0     & \cdots       &0\\
\vdots &      & \vdots&\vdots            &  &   \vdots  &    &\vdots&  \vdots     &&\vdots \\
*      &  \cdots    & *     &*          & \cdots       &*    &    \cdots            &*     &*  & \cdots     &1\\
\end{array}\right),\begin{array}{rrrrrrrrrr}
\\
\\
h+1  \\
\\
k~~\\
k+1\\
\\
k+t\\
\\
\\
\end{array}\\
\end{align*}

Now, note that $h+1<k+1$ and $\eta\neq 0$, we can do the following column transforms  for the matrix $P\big(h+1,(k+t)(\eta)\big)GH$ to make the element in the $(h+1)$th row and the $j$th column $(j=h+1,\ldots,k+t-1)$ equal to zero. 

$\bullet$ By replacing the $(h+1)$th column by the sum of $-\eta^{-1}(1+\eta l_{k+t,h+1})$ times the $(k+t)$th  column  and  the $(h+1)$th column.

$\bullet$ For any $i=h+2,\ldots,k+t-1$, by replacing the $i$th column by the sum of $-l_{k+t,i}$ times the $(k+t)$th  column  and the  $i$th column.

The result of above column transforms can be expressed as 

\begin{align*}	&P\big(h+1,(k+t)(\eta)\big)GHT_1\\
&\quad\begin{array}{cccccccccccc} 1 &  & h+1&~k+1&&k+t& & &\end{array}\\
=& \left( 
\begin{array}{cccccccccccc} 	
1      &\ldots&  0            &\cdots & 0 &\ldots   &0                 &\cdots &0 \\
\vdots &      & \vdots  &         &\vdots&      &\vdots         &       &\vdots \\
*      & \cdots          &0& \cdots&0&\ldots&\eta      &\cdots &0\\
\vdots &      & \vdots            &   &\vdots &         &\vdots  &  &\vdots \\	
*      & \cdots      & *              & \cdots   &1               & \cdots     &0                & \cdots      &0 \\
\vdots &      &\vdots            &     &  \vdots &       &\vdots      &&\vdots \\
*      &  \cdots     & *         & \cdots        &*     &\cdots      &1             & \cdots       &0\\
\vdots &      & \vdots            &    &   \vdots  &    &\vdots    &&\vdots \\
*      &  \cdots    & *          & \cdots        & *        &    \cdots            &*       & \cdots     &1\\
\end{array}\right),\begin{array}{rrrrrrrrrr}
\\
\\
h+1  \\
\\
k+1\\
\\
k+t\\
\\
\\
\end{array}
\end{align*}
where 
\begin{align*}
T_1=Q\big(h+1,(k+t)(-\eta^{-1}(1+\eta l_{k+t,h+1}))\big)\prod_{i=h+2}^{k+t-1}Q\big(i,(k+t)(-l_{k+t,i})\big).
\end{align*}

Next, in order to make the element in the $i$th row  $(i=h+1,\ldots,k+t-1)$ and the $(h+1)$th column equal to zero, note that the element in the $j$th row and the $j$th column is equal to $1$ $(j\neq h+1)$,  we can make the following   column transforms  for the matrix $P\big(h+1,(k+t)(\eta)\big)GH$  sequentially.

$\bullet$ For any $i=h+1,\ldots,k-1$, by replacing the $i$th column by the sum of $-l_{k,i}$ times the $k$th  column  and  the $i$th column.

$\bullet$ For any $i=h+1,\ldots,k-2$, by replacing the $i$th column by the sum of $-l_{k-1,i}$ times the $(k-1)$th  column  and the $i$th column.

~~$\cdots$$\cdots$$\cdots$

~~$\cdots$$\cdots$$\cdots$

~~$\cdots$$\cdots$$\cdots$

$\bullet$ For any $i=h+1,h+2$, by replacing the $i$th column by the sum of $-l_{h+3,i}$ times the $(h+3)$th  column  and the $i$th column.

$\bullet$ By replacing the $(h+1)$th column by the sum of $-l_{h+2,h+1}$ times the $(h+2)$th  column  and the $(h+1)$th column.

The result of above column transforms can be expressed as 
\begin{align*}&P\big(h+1,(k+t)(\eta)\big)GHT_1T_2\\
&\quad\begin{array}{cccccccccccc} 1 &  & h+1&~ k+1&&k+t& & &\end{array}\\
=& \left( 
\begin{array}{cccccccccccc} 	
1      &\ldots    &0            &\cdots & 0 &\ldots   &0                    &\cdots &0 \\
\vdots &      &\vdots  &         &\vdots&      &\vdots         &       &\vdots \\
*      & \cdots         &0& \cdots&0&\ldots&\eta     &\cdots &0\\
\vdots &     &\vdots            &   &\vdots &         &\vdots   &  &\vdots \\	 *      & \cdots       &0            & \cdots   &0               & \cdots     &0                 & \cdots      &0 \\
\vdots &     &\vdots            &     &  \vdots &       &\vdots     &&\vdots \\
*      &  \cdots          &*         & \cdots        &*     &\cdots      &1                  & \cdots       &0\\
\vdots &      &\vdots            &    &   \vdots  &    &\vdots    &&\vdots \\
*      &  \cdots        &*          & \cdots        & *        &    \cdots            &*    & \cdots     &1\\
\end{array}\right),\begin{array}{rrrrrrrrrr}
\\
\\
h+1  \\
\\
k~~\\
\\
k+t\\
\\
\\
\end{array}
\end{align*}where  
\begin{align*}
T_2=\prod_{i=h+1}^{k-1}Q\big(i,k(-l_{k,i})\big)\cdots\prod_{i=h+1}^{h+2}Q\big(i,(h+3)(-l_{h+3,i})\big)\cdot Q\big(h+1,(h+2)(-l_{h+2,h+1})\big).
\end{align*}

Finally, by exchanging the $(h+1)$th column and the $(k+t)$th column of the matrix $P\big((h+1),(k+t)(\eta)\big)GHT_1T_2$, one has 
\begin{align*} &P\big(h+1,(k+t)(\eta)\big)GHT_1T_2T(h+1,k+t)\\
&\quad\begin{array}{cccccccccccc} 1 &  & h+1&~ k+1&&k+t& & &\end{array}\\
=& \left( 
\begin{array}{cccccccccccc} 	
1      &\ldots    &0            &\cdots & 0 &\ldots   &0                   &\cdots &~0 \\
\vdots &      &\vdots  &         &\vdots&      &\vdots        &       &~\vdots \\
*      & \cdots         &\eta& \cdots&0&\ldots&0    &\cdots &~0\\
\vdots &     &\vdots            &   &\vdots &         &\vdots  &  &\vdots \\	 *      & \cdots       &0            & \cdots   &0               & \cdots     &0                 & \cdots      &0 \\
\vdots &     &\vdots            &     &  \vdots &       &\vdots     &&\vdots \\
*      &  \cdots          &*         & \cdots        &*     &\cdots      &1                  & \cdots       &0\\
\vdots &      &\vdots            &    &   \vdots  &    &\vdots    &&\vdots \\
*      &  \cdots        &*          & \cdots        & *        &    \cdots            &*      & \cdots     &1\\
\end{array}\right),\begin{array}{rrrrrrrrrr}
\\
\\
h+1  \\
\\
k~~\\
\\
k+t\\
\\
\\
\end{array}
\end{align*}

Now by calculating the $j$th $(j=k+1,\ldots,n)$ column of $HT_1T_2T(h+1,k+t)$, we can get $H_{n-k}(\mathbf{1})$ directly. And then from $(\ref{HH1})$, the parity-check matrix for $\mathcal{C}_{k,n}(\boldsymbol{\alpha},\boldsymbol{v},t,h,\eta)$
is given by
\begin{align*}
H_{n-k}(\boldsymbol{v})=\left(\begin{matrix}{\boldsymbol{\beta}}_1,{\boldsymbol{\beta}}_2,\ldots,{\boldsymbol{\beta}}_n
\end{matrix}
\right),
\end{align*}	 
where 
\begin{align*}
{\boldsymbol{\beta}}_{j}
=\left(\begin{matrix}
&\frac{u_j}{v_j}\\
&\frac{u_j}{v_j}\alpha_j\\
&\vdots\\
&\frac{u_j}{v_j}\alpha_j^{n-(k+t+1)}\\
&\frac{u_j}{v_j}\Big(\alpha_j^{n-(h+1)}-\sum\limits_{m=1}^{k-h-1}L_{m}\alpha_j^{n-(h+1)-m}+\tilde{L}\alpha_j^{n-(k+t)}\Big)\\
&\frac{u_j}{v_j}\big(\alpha_j^{n-(k+t-1)}-L_1\alpha_j^{n-(k+t)}\big)\\
&\vdots\\
&\frac{u_j}{v_j}\big(\alpha_j^{n-(k+1)}- L_{t-1}\alpha_j^{n-(k+t)}\big)\\
\end{matrix}\right)_{(n-k)\times 1}
\end{align*}	
and\begin{align*}
\tilde{L}=\sum\limits_{m=1}^{k-h-1}L_{m} L_{k+t-h-1-m}-\eta^{-1}(1+\eta L_{k+t-h-1}).
\end{align*}
 $\hfill\Box$\\


\section{Self-dual TGRS codes}	
\subsection{A sufficient and necessary condition for the TGRS code to be self-dual}
 Before giving a sufficient and necessary condition for $\mathcal{C}_{k,n}(\boldsymbol{\alpha},\boldsymbol{v},t,h,\eta)$$(h\ge t)$ to be self-dual, we need the following lemma.
 	\begin{lemma}\label{la} For any positive integers $m$ and $s$ with $s\ge 2$, let  $u_i=\prod\limits_{j=1,j\neq i}^{n}(\alpha_i-\alpha_{j})^{-1}$, $L_m=\sum\limits_{l=1}^{n}u_l\alpha_{l}^{n-1+m}$ and 
 		$\sigma_i=(-1)^{n-i}\sum\limits_{1\le l_1<\cdots<l_{n-i}\le n}\alpha_{l_1}\cdots\alpha_{l_{n-i}}~(0\le i\le n-1).$ Then the following two assertions hold.\\
 		
 		$(1)$ $L_1=\cdots=L_{s-1}=0$ if and only if $\sigma_{n-1}=\cdots=\sigma_{n-(s-1)}=0$.
 		
 		$(2)$ If $L_1=\cdots=L_{s-1}=0$, then $L_{2s-1}=-\sigma_{n-(2s-1)}$.
 \end{lemma}
 
 {\bf{Proof}}. $(1)$~Let $P(x)=\prod\limits_{i=1}^{n}(x-\alpha_{i})=x^{n}+\sum\limits_{i=0}^{n-1}\sigma_ix^i$, note that
 $P(\alpha_i)=0~(i=1,\ldots,n)$, thua we have $\alpha_{i}^{n}=-\sum\limits_{j=0}^{n-1}\sigma_j\alpha_i^j$. Now by Lemma \ref{l1}, for any $1\le m\le n$,  it holds that
 \begin{align}\label{lm}
 \begin{aligned}
 L_m&=-\sum_{j=1}^{n-1}\sigma_j\sum_{l=1}^{n}u_l\alpha_{l}^{j+m-1}
 =-\sum_{j=n-m}^{n-1}\sigma_j\sum_{l=1}^{n}u_l\alpha_{l}^{j+m-1}=-\sigma_{n-m}-\sum_{j=1}^{m-1}\sigma_{n-m+j}L_j.
 \end{aligned}
 \end{align}
 If $L_1=\cdots=L_{s-1}=0$, then $(\ref{lm})$ yields that
 \begin{align*}
 L_i=-\sigma_{n-i}-\sum_{j=1}^{i-1}\sigma_{n-i+j}L_j=-\sigma_{n-i}~(1\le i\le s-1),  
 \end{align*}thus $\sigma_{n-1}=\cdots=\sigma_{n-(s-1)}=0$.
 Conversely, if  $a_{n-1}=\cdots=a_{n-(s-1)}=0$, then $(\ref{lm})$ yields $L_1=\cdots=L_{s-1}=0$  directly.
 
$(2)$ If $L_1=\cdots=L_{s-1}=0$, by $(1)$, we can get
 \begin{align*}
 L_{2s-1}&=-\sigma_{n-(2s-1)}-\sum_{j=1}^{2s-2}\sigma_{n-(2s-1)+j}L_j\\
 &=-\sigma_{n-(2s-1)}-\sum_{j=1}^{s-1}\sigma_{n-(2s-1)+j}L_j-\sum_{j=1}^{s-1}\sigma_{n-s+j}L_{s-1+j}\\
 &=-\sigma_{n-(2s-1)}.
 \end{align*}
  $\hfill\Box$ 
 
\begin{theorem}\label{t1}
 Let $n$, $k$, $t$ and $h$ be non-negative integers with $n=2k$, $k\ge 4$, $t\ge 1$ and $ k-1+t\le n$. For $\eta\in\mathbb{F}_q^{*}$,  $\mathcal{C}_{k,n}(\boldsymbol{\alpha},\boldsymbol{v},t,h,\eta)$ is self-dual if the following three conditions hold.
	
	$(1)$ $h+t=k$; 
	
	$(2)$  there exists some $\lambda\in \mathbb{F}_q^{*}$ such that $\frac{v_{i}^2}{u_i}=\lambda$ $( 1\le i \le n)$;
	
	$(3)$ $\sigma_{n-m}=0~(m=1,\ldots,t-1)$ and $2-\eta \sigma _{n-(2t-1)}=0$.
	
 \noindent Conversely, if  $h\ge t$ and $\mathcal{C}_{k,n}(\boldsymbol{\alpha},\boldsymbol{v},t,h,\eta)$ is self-dual, then $(1)$-$(3)$ hold.
\end{theorem}

{\bf{Proof}}.~In the case $h=k-1$, the proof is the same as that for Theorem $2.8$ \cite{t0}. Now, we assume $h\le k-2$. It is easy to check that if conditions $(1)$-$(3)$ hold, then $\mathcal{C}_{k,n}(\boldsymbol{\alpha},\boldsymbol{v},t,h,\eta)$ is self-dual.

Under the conditions $h\ge t$ and $\mathcal{C}_{k,n}(\boldsymbol{\alpha},\boldsymbol{v},t,h,\eta)$ is self-dual, denote
\begin{align*}
G_k(\boldsymbol{v})=\left(\begin{matrix}
\mathbf{g}_0~\\
\mathbf{g}_1~\\
\vdots~\\
\mathbf{g}_{k-1}~\\
\end{matrix}\right)\qquad\text{and}\qquad
H_{n-k}(\boldsymbol{v})=\left(\begin{matrix}
\mathbf{h}_0~\\
\mathbf{h}_1~\\
\vdots~\\
\mathbf{h}_{k-1}~\\
\end{matrix}\right).
\end{align*}
Now we show that $(1)$-$(3)$ hold as follows. 

$(1)$ If $h+t>k$, namely $k-1+t>n-h-1$, since $\mathbf{g}_h\in \mathrm{Span}\{\mathbf{h}_0,\ldots,\mathbf{h}_{k-1}\}$, we know that  there exist not all zero elements $b_{0},b_{1},\ldots,b_{k-1+t}$ in $\mathbb{F}_q$ such that $\alpha_{1},\ldots,\alpha_{n}$ are roots of the polynomial $g(x)=\sum\limits_{i=0}^{k-1+t}b_ix^i$, which contradicts with  $\deg g(x)<n$, thus $h+t\le k$. Similarly, we can prove that $h+t\ge k$ by $\mathbf{h}_h\in \mathrm{Span}\{\mathbf{g}_0,\ldots,\mathbf{g}_{k-1}\}$. Thus $h+t=k$. 

$(2)$ For any $i\in\{0,\ldots,k-1\}$, we have $\mathbf{g}_i\in\mathrm{Span}  \{\mathbf{h}_{0},\ldots,\mathbf{h}_{k-1}\}$. Thus there exists a vector $(a_{i,0},a_{i,1},\ldots,a_{i,k-1}) \in \mathbb{F}_q^{k}\backslash\{(0,\ldots,0)\}$ such that $$\mathbf{g}_i=(a_{i,0},a_{i,1},\ldots,a_{i,k-1})H_{n-k}(\boldsymbol{v}).$$ Let
\begin{align*}
f_i(x)
=&a_{i,0}+a_{i,1}x+\cdots+a_{i,h-1}x^{h-1}\\
&+a_{i,h}\bigg(x^{k+t-1}-\sum\limits_{m=1}^{t-1}L_{m}x^{k+t-1-m}+\Big(\sum\limits_{m=1}^{k-h-1}L_{m} L_{k+t-h-1-m}-\eta^{-1}(1+\eta L_{k+t-h-1})\Big)x^{h}\bigg)\\
&+a_{i,h+1}\big(x^{h+1}-L_{1}x^{h}\big)+\ldots
+a_{i,k-1}\big(x^{k-1}-L_{t-1}x^{h}\big),
\end{align*}
 Then for any $i\in\{0,\ldots,k-1\}$,
\begin{align*}
\frac{v_j^{2}}{u_j}=f_0(\alpha_j),~~  \frac{v_j^{2}}{u_j}\big(\alpha_{j}^{h}+\eta\alpha_{j}^{k-1+t}\big)=f_{h}(\alpha_j)~~\text{and}~~ \frac{v_j^{2}}{u_j}\alpha_{j}^{i}=f_{i}(\alpha_j) (i\neq h),
\end{align*}  which leads
\begin{align}\label{h}
f_0(\alpha_j)\big(\alpha_{j}^{h}+\eta\alpha_{j}^{k-1+t}\big)-f_{h}(\alpha_j)=0
\end{align}
and  \begin{align}
f_0(\alpha_j)\alpha_{j}^{i}-f_{i}(\alpha_j)=0  \quad (i\neq h).
\end{align}
Note that $\alpha_{1},\ldots,\alpha_{n}$ are different roots of $f_0(x)x^{h-1}-f_{h-1}(x)$ and 
 $$\mathrm{\text{deg}}\big(f_0(x)x^{h-1}-f_{h-1}(x)\big)\le n-2,$$
 we have  $f_0(x)x^{h-1}-f_{h-1}(x)=0$, it leads $a_{0,h}=0$. Thus
\begin{align}\label{ff}\begin{aligned}
f_0(x)=&a_{0,0}+a_{0,1}x+\cdots+a_{0,h-1}x^{h-1}\\
&+a_{0,h+1}\big(x^{h+1}-L_{1}x^{h}\big)+\ldots
+a_{0,k-1}\big(x^{k-1}-L_{t-1}x^{h}\big).
\end{aligned}
\end{align} 
Now by $\mathrm{\text{deg}}\big(f_0(x)x^{k-1}-f_{k-1}(x)\big)\le 2k-2=n-2$ and $\alpha_{1},\ldots,\alpha_{n}$ are different roots of $f_0(x)x^{k-1}-f_{k-1}(x)$, one has $f_0(x)x^{k-1}-f_{k-1}(x)=0$. Since $\deg f_{k-1}(x)\le k+t-1$, we can get $\deg f_0(x)\le t$, which implies that 
\begin{align}\label{k-1}
a_{0,j}=0\quad (t+1\le j\le k-1).
\end{align}
By the assumption $h\ge t$, one has $(k-1+t)+(k-h)=n+t-h-1<n$, thus
\begin{align*}
\mathrm{deg}\Big(f_0(x)\big(x^{h}+\eta x^{k-1+t}\big)-f_{h}(x)\Big)\le n-1.
\end{align*} Since $\alpha_{1},\ldots,\alpha_{n}$ are different roots of $f_0(x)\big(x^{h}+\eta x^{k-1+t}\big)-f_{h}(x)$, one has
\begin{align}
f_0(x)\big(x^{h}+\eta x^{k-1+t}\big)-f_{h}(x)=0,
\end{align}	
it means that
\begin{align}\label{f0}
f_0(x)=a_{0,0}\neq 0,
\end{align}
thus $(2)$ holds.

$(3)$ It follows from $(\ref{h})$ and $(\ref{f0})$ that
\begin{align*}
a_{0,0}\big(x^{h}+\eta x^{k-1+t}\big)-f_{h}(x)=0,
\end{align*}
which means
\begin{align*}
&a_{0,0}\big(x^{h}+\eta x^{k+t-1}\big)\\
&-a_{h,h}\bigg(x^{k+t-1}-\sum\limits_{m=1}^{t-1}L_{m}x^{k+t-1-m}+\Big(\sum\limits_{m=1}^{k-h-1}L_{m} L_{k+t-h-1-m}-\eta^{-1}(1+\eta L_{k+t-h-1})\Big)x^{h}\bigg)\\&-a_{h,h+1}\big(x^{h+1}-L_{1}x^{h}\big)-\cdots
-a_{h,k-1}\big(x^{k-1}-L_{t-1}x^{h}\big)=0,
\end{align*}
namely,
\begin{align*}
&\Big(\eta a_{0,0}-a_{h,h}\Big) x^{k+t-1}+a_{h,h}\sum\limits_{m=1}^{t-1}L_{m}x^{k+t-1-m}-\sum_{j=1}^{t-1}a_{h,h+j}x^{h+j}\\
&\bigg(a_{0,0}-a_{h,h}\Big(\sum\limits_{m=1}^{t-1}L_{m} L_{2t-1-m}-\eta^{-1}(1+\eta L_{2t-1})\Big)+\sum_{j=1}^{t-1}a_{h,h+j}L_j\bigg)x^{h}=0,
\end{align*}
and so
\begin{align}\label{E1}
\begin{cases}
\eta a_{0,0}-a_{h,h}=0;\\
a_{h,h}L_m=0~~(m=1,\ldots,t-1);\\
a_{h,h+j}=0~~(j=1,\ldots,t-1);\\
a_{0,0}-a_{h,h}\Big(\sum\limits_{m=1}^{t-1}L_{m} L_{2t-1-m}-\eta^{-1}(1+\eta L_{2t-1})\Big)+\sum\limits_{j=1}^{t-1}a_{h,h+j}L_j=0.
\end{cases}
\end{align}
Note that $a_{h,h}=\eta a_{0,0}\neq 0$, thus
\begin{align}\label{E2}  
\begin{cases}
L_m=0~~(m=1,\ldots,t-1);\\
2+\eta L_{2t-1}=0.   
\end{cases}
\end{align}
Now by $(\ref{E2})$ and Lemma $\ref{la}$, $(3)$ holds. $\hfill\Box$

\begin{remark}
	If $2\mid q$, then Theorem \ref{t1} $(3)$ can be replaced by $$\sigma_{n-(2t-1)}=\sigma_{n-m}=0~(m=1,\ldots,t-1).$$
\end{remark}
\subsection{The existence for self-dual TGRS codes over $\mathbb{F}_q$ $(2\mid q)$}
\begin{theorem}\label{t21}
	Let  $t$, $k$, $s$ and $m$ be positive integers with $s\mid m$ and $t\le 2^{s-1}-2$. Let  $\mathbb{F}_{2^{s}}=\{\alpha_1,\ldots,\alpha_{2^{s}}\}$ be the subfield of $\mathbb{F}_{2^{m}}$ and $\eta\in\mathbb{F}_{2^m}^{*}$.  If $\boldsymbol{\alpha}=(\alpha_1,\ldots,\alpha_{2^{s}})$ and $\texorpdfstring{\bm{v}}~=(v_1,\ldots,v_{2^{s}})$ with $v_i=\sum_{j=1,j\neq i}^{2^{s}}(\alpha_i-\alpha_j)^{-2^{m-1}}$ $(i=1,\ldots,2^{s})$, then
	$\mathcal{C}_{2^{s-1},2^s}(\texorpdfstring{\bm{\alpha}}~,\texorpdfstring{\bm{v}}~,t,2^{s-1}-t,\eta)$ is a self-dual $m$-MDS code with $m\le t$. 
\end{theorem}

{\bf Proof.} It is easy to see that $\prod\limits_{i=1}^{2^s}(x-\alpha_i)=x^{2^s}-x$, thus 
$$\sigma_j=(-1)^{n-j}\sum\limits_{1\le l_1<\cdots<l_{n-j}\le n}\alpha_{l_1}\cdots\alpha_{l_{n-j}}=0~~(j=2,\ldots,2^{s}-1).$$
Now by $v_i^2=u_i$ $(i=1,\ldots,2^s)$ and  Theorem $\ref{t1}$, 
$\mathcal{C}_{2^{s-1},2^s}(\texorpdfstring{\bm{\alpha}}~,\texorpdfstring{\bm{v}}~,t,2^{s-1}-t,\eta)$ is self-dual. 
And then by Lemma $\ref{lh}$, we complete the proof. $\hfill\Box$\\

By Lemma \ref{lmds} and Theorem \ref{t21}, we have the following corollary directly.
\begin{corollary}
By taking $\eta\in\mathbb{F}_{2^m}^{*}\backslash\mathbb{F}_{2^s}$ in Theorem \ref{t21}, then $\mathcal{C}_{2^{s-1},2^s}(\texorpdfstring{\bm{\alpha}}~,\texorpdfstring{\bm{v}}~,t,2^{s-1}-t,\eta)$ is a self-dual MDS  code.
\end{corollary}
\begin{theorem}\label{t22}
	For positive integers $m$ and $l$ with $l\le m$, let $w_1,\ldots,w_m$ be a basis of $\mathbb{F}_{2^{m}}$ over $\mathbb{F}_2$, denote $$\{\alpha_1,\ldots,\alpha_{2^{l}}\}=\{b_1w_1+\cdots+b_lw_l~|~b_i\in\mathbb{F}_2~(i=1,\ldots,l)\}.$$ Then for any  positive integer $k\ge 2^{l-1}$, $\eta\in \mathbb{F}_{2^m}^{*}$, $\boldsymbol{\alpha}=(\alpha_1,\ldots,\alpha_{2k})$ and $\texorpdfstring{\bm{v}}~=(v_1,\ldots,v_{2k})$ with $v_i=\sum\limits_{j=1,j\neq i}^{2k}(\alpha_i-\alpha_j)^{-2^{m-1}}$ $(i=1,\ldots,2k)$, $\mathcal{C}_{k,2k}(\texorpdfstring{\bm{\alpha}}~,\texorpdfstring{\bm{v}}~,1,k-1,\eta)$ is self-dual  if one of the following conditions holds.
	
 $(1)$  $k=2^{l-1}$;
	
$(2)$  $l_1$ is odd with $3\le l_1 \le m-l$ and $k=\frac{2^{l}+l_1+1}{2}$, $\alpha_{2^l+j}=w_{l+j} (j=1,\ldots,l_1)$ and $\alpha_{2^l+l_1+1}=\sum\limits_{j=1}^{l_1}w_{l+j}$;
	
$(3)$  $l_1$ is even with $4\le l_1 \le m-l$ and $k=\frac{2^{l}+l_1}{2}$, $\alpha_{2^l+j}=w_{l+j}+w_{l+j+1} (j=1,\ldots,l_1-1)$ and $\alpha_{2^l+l_1}=w_{l+1}+w_{l+l_1}$. \\
\end{theorem}

{\bf Proof}. ~If one of the conditions $(1)$-$(3)$ holds, it is easy to verify that $\sum_{i=1}^{2k}\alpha_i=0$ and $v_j^2=u_j$ $(j=1,\ldots,2^s)$. Then by Theorem $\ref{t1}$, $\mathcal{C}_{k,2k}(\texorpdfstring{\bm{\alpha}}~,\texorpdfstring{\bm{v}}~,1,k-1,\eta)$ is self-dual.$\hfill\Box$\\

By Lemma \ref{l} and Theorem $\ref{t22}$, we have the following corollary directly.
\begin{corollary}
	In Theorem \ref{t22}, if $\eta\notin S_l(1,\boldsymbol{\alpha})$, then  $\mathcal{C}_{k,2k}(\texorpdfstring{\bm{\alpha}}~,\texorpdfstring{\bm{v}}~,1,k-1,\eta)$  is a self-dual MDS. Otherwise, $\mathcal{C}_{k,2k}(\texorpdfstring{\bm{\alpha}}~,\texorpdfstring{\bm{v}}~,1,k-1,\eta)$  is a self-dual NMDS  code.
\end{corollary}
\begin{theorem}\label{tt2}
	Let $s$, $\lambda$, $l$, $t$ be positive integers with $t<l-1$, $q=2^s$, $q_1=2^{\lambda}$ and $\eta\in\mathbb{F}_q^{*}$. Assume that $\mathbb{F}_q$ is the splitting field of $A(x)=x^{2l}+bx+c~(b,c\in\mathbb{F}_{q_1}^{*})$ over $\mathbb{F}_{q_1}$ and  $\alpha_{1},\ldots,\alpha_{2l}$ are all roots of $A(x)$ over $\mathbb{F}_q$. If $\boldsymbol{\alpha}=(\alpha_1,\ldots,\alpha_{2l})$, then $C_{l,2l}(\boldsymbol{\alpha},\mathbf{1},t,l-t,\eta)$ is a self-dual $m$-MDS code with $m\le t$.
\end{theorem}

{\bf {Proof}}.
On the one hand, since $A(x)=x^{2l}+bx+c$, we have $A^{'}(x)=b\neq 0$, and then $\gcd(A(x),A^{'}(x))=1$, namely, $\alpha_{i}\neq\alpha_{j}$ $(i\neq j)$. On the other hand, by the assumptions, we have $A(x)=\prod\limits_{i=1}^{2l}(x-\alpha_i)$, so $\sigma_{2l-(2t-1)}=\sigma_{2l-i}=0$ $(i=1,\ldots,{2t-2})$ and $A^{'}(x)=\sum\limits_{i=1}^{2l}\prod\limits_{j\neq i,j=1}^{2l}(x-\alpha_j)$, which means
\begin{align*}
u_i=A^{'}(\alpha_i)^{-1}=-{b}^{-1}\neq 0.
\end{align*}
Now by Theorem \ref{t1} and Lemma \ref{lh}, we complete the proof. $\hfill\Box$\\

By Theorem \ref{tt2} and Lemmas \ref{l}-\ref{l2}, we have the following two corollaries.

\begin{corollary}
	By taking $t=1$ in Theorem \ref{tt2},  if $\eta\notin S_l(1,\boldsymbol{\alpha})$, then $\mathcal{C}_{l,2l}(\texorpdfstring{\bm{\alpha}}~,\texorpdfstring{\bm{1}}~,1,l-1,\eta)$ is a self-dual MDS code. Otherwise, $\mathcal{C}_{l,2l}(\texorpdfstring{\bm{\alpha}}~,\texorpdfstring{\bm{1}}~,1,l-1,\eta)$ is a self-dual NMDS code.
\end{corollary}

\begin{corollary}
	By taking $t=2$ in Theorem \ref{tt2}, the following assertions hold.

	$(1)$ If $\eta\notin S_l(2,\boldsymbol{\alpha})$, then $\mathcal{C}_{l,2l}(\texorpdfstring{\bm{\alpha}}~,\texorpdfstring{\bm{1}}~,2,l-2,\eta)$ is a self-dual MDS code;

	$(2)$ If $\eta^{-1}\in S_l(2,\boldsymbol{\alpha})\backslash \tilde{S}_l(2,\boldsymbol{\alpha})$, then $\mathcal{C}_{l,2l}(\texorpdfstring{\bm{\alpha}}~,\texorpdfstring{\bm{1}}~,2,l-2,\eta)$ is a self-dual NMDS code;

	$(3)$ If $\eta^{-1}\in \tilde{S}_l(2,\boldsymbol{\alpha})\cap S_l(2,\boldsymbol{\alpha})$, then $\mathcal{C}_{l,2l}(\texorpdfstring{\bm{\alpha}}~,\texorpdfstring{\bm{1}}~,2,l-2,\eta)$ is a self-dual $2$-MDS code.
\end{corollary}

\subsection{The existence for a self-dual TGRS code over $\mathbb{F}_q$ $(2\nmid q)$}

\begin{theorem}\label{t2}	Let $s$, $\lambda$, $l$, $t$  be positive integers, $p$ an odd prime with $p\nmid(2t-1)$ and $lp\ge t$, $q=p^s$ and $q_1=p^{\lambda}$. Assume that $\mathbb{F}_q$ is the splitting field of  $m(x)=x^{2lp}+bx^{2lp-(2t-1)}+c~(b,c\in\mathbb{F}_{q_1}^{*})$ over $\mathbb{F}_{q_1}$, $\alpha_{1},\ldots,\alpha_{2lp}$ are all roots of $m(x)$ and $v_i=\alpha_i^{t-lp}$ $(i=1,\ldots,2lp)$. Let  $\boldsymbol{\alpha}=(\alpha_1,\ldots,\alpha_{2lp})$ and $\texorpdfstring{\bm{v}}~=(v_1,\dots,v_{2lp})$ and $\eta=2{b^{-1}}$, then $C_{lp,2lp}(\boldsymbol{\alpha},\texorpdfstring{\bm{v}}~,t,lp-t,\eta)$ is a self-dual $m$-MDS code with $m\le t$.
\end{theorem}

{\bf {Proof}}. On the one hand, since  $m(x)=x^{2lp}+bx^{2lp-(2t-1)}+c~(b,c\in\mathbb{F}_{q_1}^{*})$, we have $m^{'}(x)=-(2t-1)bx^{2lp-2t}$, and then $\gcd(m(x),m^{'}(x))=1$, namely,  $\alpha_{i}\neq\alpha_{j}$ $(i\neq j)$. On the other hand, by the assumptions, we have $m(x)=\prod\limits_{i=1}^{2lp}(x-\alpha_i)$, so $\sigma_{2lp-i}=0$ $(i=1,\ldots,{2t-2})$, $2\!-\!\eta \sigma_{2lp-(2t-1)}=0$ and  
$m^{'}(x)=\sum\limits_{i=1}^{2lp}\prod\limits_{j\neq i,j=1}^{2lp}(x-\alpha_j)$, which means
\begin{align*}
u_i=m^{'}(\alpha_i)^{-1}=-((2t-1)b)^{-1}\alpha_i^{2t-2lp}=-((2t-1)b)^{-1}v_i^2\neq 0.
\end{align*}
By Theorem \ref{t1} and Lemma \ref{lh}, we complete the proof. $\hfill\Box$\\

By Theorem \ref{t2} and Lemma \ref{l2}, we have the following corollary.

\begin{corollary}
	By taking $t=2$ in Theorem \ref{t2}, the following assertions hold.

	$(1)$ If $\eta\notin S_{lp}(2,\boldsymbol{\alpha})$, then $\mathcal{C}_{lp,2lp}(\texorpdfstring{\bm{\alpha}}~,\texorpdfstring{\bm{v}}~,2,lp-2,\eta)$ is a  self-dual MDS code.

	$(2)$ If $\eta\in S_{lp}(2,\boldsymbol{\alpha})\backslash \tilde{S}_{lp}(2,\boldsymbol{\alpha})$, then $\mathcal{C}_{lp,2lp}(\texorpdfstring{\bm{\alpha}}~,\texorpdfstring{\bm{v}}~,2,lp-2,\eta)$ is a self-dual NMDS code.
	
	$(3)$ If $\eta\in S_{lp}(2,\boldsymbol{\alpha})\cap \tilde{S}_{lp}(2,\boldsymbol{\alpha})$, then $\mathcal{C}_{lp,2lp}(\texorpdfstring{\bm{\alpha}}~,\texorpdfstring{\bm{v}}~,2,lp-2,\eta)$ is a  self-dual $2$-MDS code.
\end{corollary}

\begin{theorem}\label{tp}
Let $p$ be an odd prime,  $t$, $s$ and $m$ be positive integers with $s\mid \frac{m}{2}$, $\beta\in\mathbb{F}_{p^m}^{*}\backslash\mathbb{F}_{p^s}$ and $\eta=-2\beta^{-1}$. Denote  $\{\alpha_1,\ldots,\alpha_{p^{s}-1}\}=\{\beta+a|a\in\mathbb{F}_{p^{s}}^{*}\}$, then $\mathcal{C}_{\frac{p^s-1}{2},p^s-1}(\texorpdfstring{\bm{\alpha}}~,\texorpdfstring{\bm{v}}~,1,\frac{p^s-1}{2}-1,\eta)$ is self-dual, where $\boldsymbol{\alpha}=(\alpha_1,\ldots,\alpha_{p^{s}-1})$ and $\texorpdfstring{\bm{v}}~=(v_1,\ldots,v_{p^{s}-1})$ with $v_i^2=\alpha_i^{-1}$ $(i=1,\ldots,p^{s})$.
\end{theorem}

{\bf Proof}.~By calculating directly, one has $$\sum_{j=1}^{p^s-1}\alpha_i=\sum_{a\in\mathbb{F}_{p^s}^{*}}(\beta+a)=-\beta.$$ Now by $\eta=-2\beta^{-1}$, we have	$2-\eta\sum\limits_{j=1}^{p^s-1}\alpha_i=0$. Furthermore,
\begin{align*}
\sum_{j=1,j\neq i}^{p^{s}-1}(\alpha_i-\alpha_j)^{-1}=\alpha_i^{-1}~(i=1,\ldots,p^{s}-1).
\end{align*}
By $s\mid \frac{m}{2}$, we know that $-\alpha_i$ $(i=1,\ldots,p^{s}-1)$ is a square element in $\mathbb{F}_{p^m}$, and then,
	$$v_i^2=\sum_{j=1,j\neq i}^{p^{s}-1}(\alpha_i-\alpha_j)^{-1}.$$
By Theorem \ref{t1}, $\mathcal{C}_{\frac{p^s-1}{2},p^s-1}(\texorpdfstring{\bm{\alpha}}~,\texorpdfstring{\bm{v}}~,1,\frac{p^s-1}{2}-1,\eta)$ is self-dual. $\hfill\Box$\\

By Theorem \ref{t2} and Lemma \ref{l}, we have the following corollary directly.
\begin{corollary}
	In Theorem \ref{tp}, if $\eta\notin S_l(1,\boldsymbol{\alpha})$, then  $\mathcal{C}_{\frac{p^s-1}{2},p^s-1}(\texorpdfstring{\bm{\alpha}}~,\texorpdfstring{\bm{v}}~,1,\frac{p^s-1}{2}-1,\eta)$  is a self-dual MDS code. Otherwise, $\mathcal{C}_{\frac{p^s-1}{2},p^s-1}(\texorpdfstring{\bm{\alpha}}~,\texorpdfstring{\bm{v}}~,1,\frac{p^s-1}{2}-1,\eta)$  is a self-dual NMDS  code. 
\end{corollary}

 
 
\section{Conclusions }

 In \cite{t0}, the authors gave the parity-check matrix for $\mathcal{C}_{k,n}(\boldsymbol{\alpha},\boldsymbol{v},1,k-1,\eta)$, and then obtained some self-dual MDS or NMDS codes.
 In our works,  we give the following main results for $\mathcal{C}_{k,n}(\boldsymbol{\alpha},\boldsymbol{v},t,h,\eta)$, which extends the main results in \cite{t0}.

$(1)$	The parity-check matrix for $\mathcal{C}_{k,n}(\boldsymbol{\alpha},\boldsymbol{v},t,h,\eta)$.

$(2)$	A sufficient and necessary condition for  $\mathcal{C}_{k,n}(\boldsymbol{\alpha},\boldsymbol{v},t,h,\eta)$ $(h\ge t)$ to be self-dual;

$(3)$   Construction for several classes of self-dual $m$-MDS codes with $m\le t$ from $\mathcal{C}_{k,n}(\boldsymbol{\alpha},\boldsymbol{v},t,h,\eta)$.\\

  \qquad\\
 \noindent{\large \bf Acknowledgement~} This research was supported by the National Science Foundation of China (12071321).\\
 \qquad\\
 
 \noindent{\large\bf Conflict of interest~} The authors have no conflicts of interest to declare that are relevant to the content of this
 article.\\
 
 \noindent{\large\bf Data availibility~} Not applicable.\\
 
 \noindent{\large\bf Code Availability~} Not applicable.\\

 \noindent{\large\bf Ethical approval~}Not applicable.\\
 
 \noindent{\large\bf Consent to participate~}Not applicable.\\
 
 \noindent{\large\bf Consent for publication~}Not applicable.\\

\end{document}